\documentclass[twocolumn,showpacs,preprintnumbers,amsmath,amssymb,superscriptaddress]{revtex4}%
\usepackage{graphicx}
\usepackage{color}
\usepackage{epstopdf}
\usepackage{amsmath}
\usepackage{amssymb}
\usepackage{dcolumn}
\usepackage{mathrsfs}
\usepackage{bm}

\begin{document}
	
	\title{Simulating the Kibble-Zurek mechanism of the Ising model with a
		superconducting qubit system}
	\author{Ming Gong}
	\thanks{These authors contributed equally to this work.}
	\affiliation{National Laboratory of Solid State Microstructures, School of Physics,
		Nanjing University, Nanjing 210093, China}
	\affiliation{Department of Physics and Astronomy, University of Kansas, Lawrence, KS
		66045, USA}
	\author{Xueda Wen}
	\thanks{These authors contributed equally to this work.}
	\affiliation{Department of Physics, University of Illinois at Urbana-Champaign, Urbana,
		IL 61801, USA, }
	\author{Guozhu Sun}
	\thanks{These authors contributed equally to this work.}
	\affiliation{Research Institute of Superconductor Electronics, School of Electronic\\
		Science and Engineering, Nanjing University, Nanjing 210093, China,}
	\affiliation{Synergetic Innovation Center of Quantum Information and Quantum Physics,
		University of Science and Technology of China, Hefei, Anhui 230026, China,}
	\author{Dan-Wei Zhang}
	\affiliation{Guangdong Provincial Key Laboratory of Quantum Engineering and Quantum
		Materials, SPTE, South China Normal University, Guangzhou 510006, China}
	\author{Dong Lan}
	\affiliation{National Laboratory of Solid State Microstructures, School of Physics,
		Nanjing University, Nanjing 210093, China}
	\author{Yu Zhou}
	\affiliation{Research Institute of Superconductor Electronics, School of Electronic\\
		Science and Engineering, Nanjing University, Nanjing 210093, China,}
	\author{Yunyi Fan}
	\affiliation{Research Institute of Superconductor Electronics, School of Electronic\\
		Science and Engineering, Nanjing University, Nanjing 210093, China,}
	\author{Yuhao Liu}
	\affiliation{National Laboratory of Solid State Microstructures, School of Physics,
		Nanjing University, Nanjing 210093, China}
	\author{Xinsheng Tan}
	\affiliation{National Laboratory of Solid State Microstructures, School of Physics,
		Nanjing University, Nanjing 210093, China}
	\author{Haifeng Yu}
	\affiliation{National Laboratory of Solid State Microstructures, School of Physics,
		Nanjing University, Nanjing 210093, China}
	\affiliation{Synergetic Innovation Center of Quantum Information and Quantum Physics,
		University of Science and Technology of China, Hefei, Anhui 230026, China,}
	\author{Yang Yu}
	\email{yuyang@nju.edu.cn}
	\affiliation{National Laboratory of Solid State Microstructures, School of Physics,
		Nanjing University, Nanjing 210093, China}
	\affiliation{Synergetic Innovation Center of Quantum Information and Quantum Physics,
		University of Science and Technology of China, Hefei, Anhui 230026, China,}
	\author{Shi-Liang Zhu}
	\email{slzhu@nju.edu.cn}
	\affiliation{National Laboratory of Solid State Microstructures, School of Physics,
		Nanjing University, Nanjing 210093, China}
	\affiliation{Synergetic Innovation Center of Quantum Information and Quantum Physics,
		University of Science and Technology of China, Hefei, Anhui 230026, China,}
	\author{Siyuan Han}
	\email{han@ku.edu}
	\affiliation{Department of Physics and Astronomy, University of Kansas, Lawrence, KS
		66045, USA}
	\author{Peiheng Wu}
	\affiliation{Research Institute of Superconductor Electronics, School of Electronic\\
		Science and Engineering, Nanjing University, Nanjing 210093, China,}
	\affiliation{Synergetic Innovation Center of Quantum Information and Quantum Physics,
		University of Science and Technology of China, Hefei, Anhui 230026, China,}

\begin{abstract}
The Kibble-Zurek mechanism (KZM) predicts the density of topological defects
produced in the dynamical processes of phase transitions in systems ranging
from cosmology to condensed matter and quantum materials. The similarity
between KZM and the Landau-Zener transition (LZT), which is a standard tool
to describe the dynamics of some non-equilibrium physics in contemporary
physics, is being extensively exploited. Here we demonstrate the equivalence
between KZM in the Ising model and LZT in a superconducting qubit system. We
develop a time-resolved approach to study quantum dynamics of LZT with
nano-second resolution. By using this technique, we simulate the key
features of KZM in the Ising model with LZT, e.g., the boundary between the
adiabatic and impulse regions, the freeze-out phenomenon in the impulse
region, especially, the scaling law of the excited state population as the
square root of the quenching rate. Our results supply the experimental
evidence of the close connection between KZM and LZT, two textbook paradigms
to study the dynamics of the non-equilibrium phenomena.
\end{abstract}

\maketitle


\newpage

Non-equilibrium phenomena at avoided level crossings play an essential role
in many dynamical processes throughout physics and chemistry. A transition
between energy levels at the avoided crossing is known as the
Landau-Zener transition (LZT) \cite{Landau,zener1932non}, which has served over decades
as a textbook paradigm of quantum dynamics. LZT has recently been
extensively studied \cite{Shevchenko} both theoretically and experimentally
in, e.g., superconducting qubits \cite{Oliver,sillanpaa2006continuous,Tan,sun2010tunable}, spin-transistor \cite{Betthausen}
, and optical lattices \cite{Tarruell,salger2007atomic,Chen,Zhang}. On the other
hand, quantum phase transition may also relate to avoided level crossings
and it plays an important role in nature. Recently, an elegant theoretical
framework for understanding the dynamics of phase transition is provided by
the Kibble-Zurek mechanism (KZM) \cite{Kibble,kibble1980some,Zurek,zurek1996cosmological}. When the parameters of
a quantum system that drive the quantum phase transition are varied
in time causing the system to traverse the critical point, KZM
predicts that the density of the defects produced in the processes
follows a power law that scales with the square root of the speed at
which the critical point is traversed. Due to its ubiquitous nature, this
theory finds applications in a wide variety of systems ranging from
cosmology to condensed matter and quantum materials \cite{Chuang,Ulm,Pyka,Navon}.

The correspondence between LZT and KZM was first pointed out by Damski \cite{Damski,damski2006adiabatic}. 
It was shown that the dynamics of LZT can be intuitively described
in terms of KZM of the topological defect production in nonequilibrium
quantum phase transition. In order to model the dynamical process of LZT, a
widely used picture is the adiabatic-impulse approximation (AIA), which was
originally developed in KZM theory. The entire dynamical process can be
divided into three regions: the adiabatic, impulse, and adiabatic regions,
as shown in Fig. 1a. The three regions are separated by two boundaries $-v\hat{t}
$ and $v\hat{t}$, where $v$ is the quench rate and $\hat{t}$ is referred to
as the freeze out time. Based on AIA, the dynamics of topological defect
production in non-equilibrium phase transitions can be simulated with LZT.
Recently this prediction was experimentally confirmed in an optical
interferometer\cite{Xu}. However, some key features in the correspondence
between LZT and KZM, such as the freeze out time $\hat{t}$ and the
adiabatic-impulse-adiabatic regions, have not been investigated 
experimentally. Most importantly, by studying the dynamical quantum phase
transition in a quantum Ising chain, it is found that the average density of
defects scales as the square root of the quenching rate \cite{Zurek2005,Dziarmaga}. 
This universal scaling law of defect formulation as a
function of quench time, which lies at the heart of KZM, lacks
adequate experimental evidence in LZT.

\begin{figure}
	\begin{center}
		\includegraphics[width=.5\textwidth]{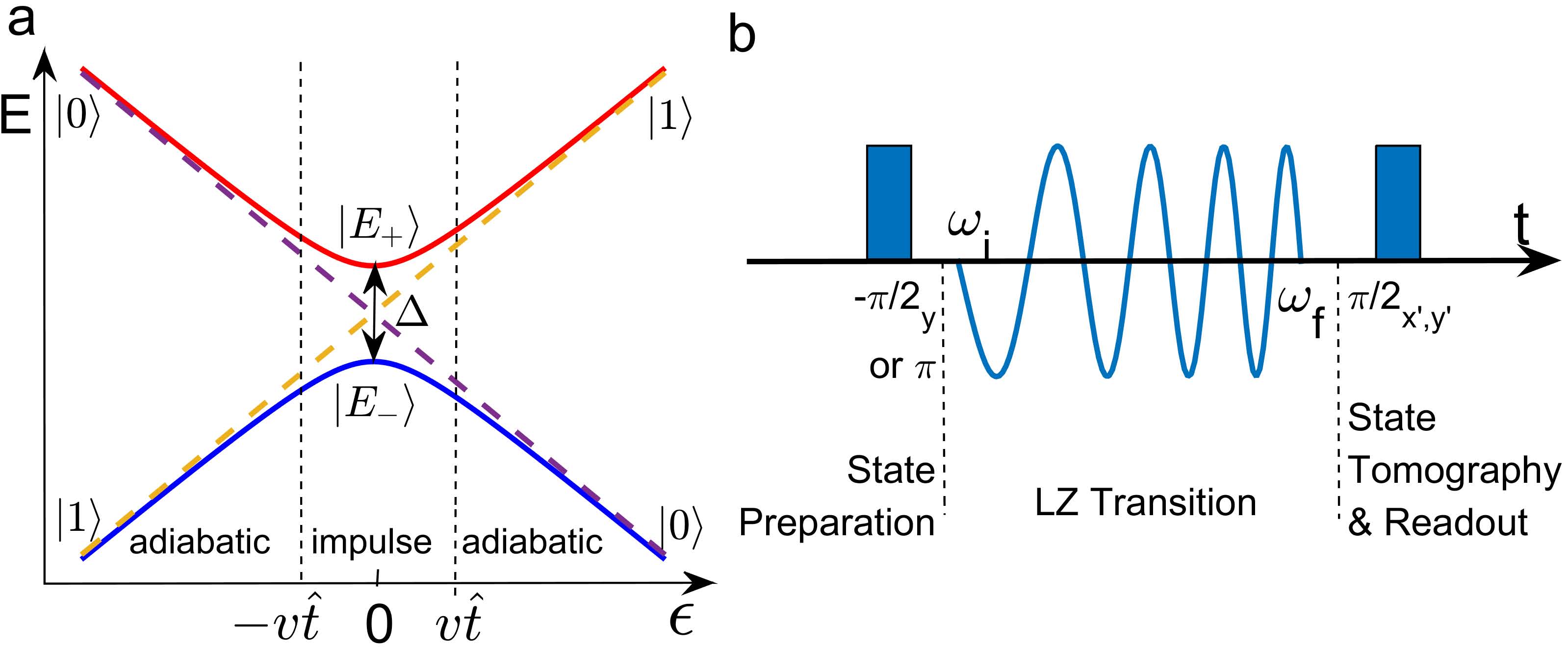}
		\caption{\textbf{Energy level avoided crossing and experimental procedure. }%
			\textbf{(a)} A typical energy structure (parameterized by time) of a
			two-level system. The diabatic states $|{}0\rangle ,$ $|{}1\rangle ,$ and
			energy eigenstates $|{}E_{\mp }\rangle $ are denoted in the plot. \textbf{(b)%
			} A schematic of time profile of the experiment consisting of three parts.
			During state preparation, a $-\protect\pi /2_{y} $ ($\protect\pi $) pulse is
			applied to prepare the qubit in $(|0\rangle +|1\rangle )/\protect\sqrt 2$ ($%
			|1\rangle) $. The Landau-Zener transition is realized by chirping the
			microwave frequency from $\protect\omega _{i}$ to $\protect\omega _{f}$. The
			final state of the qubit is obtained by state tomography. }
		\label{fig:Graph1}
	\end{center}
\end{figure}

In this paper, we use LZT in superconducting qubits to simulate KZM of
the Ising model. We develop a time-resolved method to directly investigate
the quantum dynamics of LZT in the superconducting qubit. Using state
tomography, we measure the time evolution of the population $P_{+}$ of the instantaneous positive
energy eigenstate [see Fig. 1a] for the entire LZT
process. We find that $P_{+}$ exhibits a rapid change near
the center of the avoided crossing and varies gradually 
outside this region, revealing the existence of the
adiabatic and impulse regions. Moreover, the freeze-out behavior 
predicted by KZM has also been observed, 
and the boundary between the adiabatic and impulse region predicted by AIA
is confirmed. We observe that the experimental simulated KZM of Ising model
displays the theoretically predicted Kibble-Zurek scaling law. Therefore, we
experimentally demonstrate the close connection between KZM and LZT, in
particular, the presence of Kibble-Zurek scaling behavior in LZT.

\noindent \textbf{Results}\newline
\noindent \textbf{The equivalence between KZM and LZT.} The Ising model is
regarded as one of the two prototypical models to understand quantum phase
transitions \cite{Sachdev}. After rescaling all the quantities to the
dimensionless variables, we obtain the Ising model Hamiltonian  
\begin{equation}
H_{I}=-\sum_{n=1}^{N}(g\sigma _{n}^{x}+\sigma _{n}^{z}\sigma _{n+1}^{z}),
\label{HI}
\end{equation}%
with the periodic boundary condition, where $\sigma _{n}^{x, z}$ are the Pauli-matrices operators. Here $N$ is the total number
of spins and $g$ is a dimensionless constant driving the phase transition.
The ground state of $H_{I}$ is a paramagnet for $g\gg 1$ and a ferromagnet
for $g\ll 1$, and $g=1$ corresponds to the critical point. To study the
dynamics of this model, we assume that the system evolves from time $%
t_{i}=-\infty $ to $t_{f}=0$, and takes a linear quench $g(t<0)=-t/\tau
_{Q}^{I}$, where $1/\tau _{Q}^{I}$ provides a quench rate.

By utilizing Jordan-Wigner transformation and Fourier transform, the Ising
model can be simplified as a bunch of decoupled qubits with the Hamiltonian $%
H_{I}^{\prime }=\sum_{k}\epsilon _{k}\left( \gamma _{k}^{\dag }\gamma _{k}-%
\frac{1}{2}\right) $, where the eigenenergy $\epsilon _{k}=2\sqrt{[g-\cos
	k]^{2}+\sin ^{2}k}$ and $k=\pm \frac{1}{2}\frac{2\pi }{N},\cdots ,\pm (\frac{%
	N}{2}-\frac{1}{2})\frac{2\pi }{N}$ is the pseudomomentum. Here and hereafter
we set the lattice constant $a=1$. The density of defects resulting from the
quantum quench has the expression $\mathcal{N}=\sum_{k}p_{k}/N$, where $p_{k}
$ is the excitation probability corresponding to the pseudomomentum $k$. We
consider a special case where the system undergoes slow evolution with $\tau
_{Q}^{I}\gg 1/2\pi $. Under this condition, it is safe to assume that only
long wavelength modes are excited, i.e., $k\ll \pi /4$, and then $%
p_{k}\approx |u_{k}(t_{f})|^{2}$, where $\{u_{k},v_{k}\}$ are Bogoliubov
modes governed by the following matrix equation
\begin{equation}
i\hbar \frac{d}{d\tau }\left( 
\begin{array}{c}
v_{k}  \\ 
u_{k} 
\end{array}%
\right) =\frac{1}{2}\left( 
\begin{array}{cc}
\tau \chi_{k} & 1\\ 
1 & -\tau \chi_{k}
\end{array}%
\right) \left( 
\begin{array}{c}
v_{k}\\ 
u_{k}
\end{array}%
\right) .  \label{LZIsing}
\end{equation}%
Here $\tau =4(t+\tau _{Q}^{I}\cos k)\sin k$ is the normalized time, and $%
\chi_{k}=1/4\tau _{Q}^{I}\sin ^{2}k$ is the sweeping velocity.

On the other hand, for a quantum two-level system with the diabatic basis ,
we consider the time-dependent Hamiltonian of a quantum two-level system in
the diabatic basis $|{}0\rangle $ and $|{}1\rangle $ 
\begin{equation}
H\left( t\right) =-\frac{1}{2}\left( 
\begin{array}{cc}
\epsilon (t) & \Delta {} \\ 
\Delta {} & -\epsilon (t)%
\end{array}%
\right) ,  \label{H0}
\end{equation}%
where $\Delta$ is the tunneling amplitude and $\epsilon (t)$ is the energy
difference between the two diabatic basis. We mainly consider $\epsilon
(t)=vt$ with $v$ being the speed of energy variation. The time-dependent
instantaneous energy eigenstates $|{}E_{\pm }(t)\rangle $ are 
\begin{equation}
\begin{array}{ll}
|E_{-}\left( t\right) \rangle = \cos (\theta /2) |0\rangle + \sin (\theta
/2) |1\rangle &  \\ 
|E_{+}\left( t\right) \rangle = -\sin (\theta /2) |0\rangle + \cos (\theta
/2) |1\rangle & 
\end{array}
,
\end{equation}%
where $\cos \theta =\epsilon (t)/\Omega (t)$ with $\Omega {}\left( t\right) =%
\sqrt{{\Delta {}}^{2}+\epsilon (t)^2}$. The instantaneous energy
eigenvalues of $H(t)$ are $E_{\pm }(t)=\pm \frac{1}{2}\Omega (t)$, forming
an avoided level crossing at $t=0$ with a gap $\Delta$. If the system
initially in the ground state traverses the avoided crossing, the
Landau-Zener theory gives the probability of the qubit occupying the
exciting state as $P_{LZ} \approx \exp(-\pi \Delta^2/2\hbar v)$. LZT has
received tremendous attention since quantum two-level systems, i.e., qubits,
are currently considered as the best building blocks of quantum information
processors.

With the substitution $v/\Delta^2=\chi_{k}$ and $t\Delta =\tau$, it is found that the dynamics of
the Ising model governed by Eq. (\ref{LZIsing}) is the same as LZT physics
contained in Eq. (\ref{H0}) up to a normalized tunneling amplitude $\Delta$%
\cite{Dziarmaga}. Therefore, we can use LZT to simulate KZM of the Ising
model.

A key concept of KZM is AIA. Following the arguments in Ref. \cite{Damski,damski2006adiabatic},
we consider two nontrivial schemes to relate LZT with KZM. In scheme 
$\mathbf{A}$, the system starts far away from the avoided crossing,
corresponding to $\epsilon _{i}\rightarrow -\infty $, and ends
also far away from the avoided crossing, corresponding to $\epsilon
_{f}\rightarrow \infty $. The initial state $|\psi _{0}\rangle $ is the
ground state of the Hamiltonian at time $t=-\infty $. As shown in Fig.
1a, the evolution can be divided into three regions: \
there are two adiabatic regions $t=(-\infty ,$ $-\hat{t})$ and $(%
\hat{t},$ $\infty )$, where almost no transition between the instantaneous
energy eigenstates $|{}E_{\pm }(t)\rangle $ occurs. On the
contrary, in the impulse region $[-\hat{t},\hat{t}]$ transitions
between the states $|E_{\pm }(t)\rangle $ could occur. 
Quantitatively, the boundaries separating the adiabatic
and impulse regions are determined by the freeze-out time $\hat{t}=%
\sqrt{\tau _{Q}\tau _{0}/\alpha },$ where $\alpha $ is a 
dimensionless parameter, $\tau _{Q}=\Delta /v$ sets the scale of
the quench time and $\tau _{0}=1/\Delta $, respectively. 
The finite density of topological defects $\mathcal{D}_{n}$ is
caused by non-adiabatic evolution in the impulse region $[-\hat{t},\hat{t}]$%
, which equals the occupation probability of $|{}E_{+}\rangle $.
Therefore, the density of topological defects $D_{n}$ in KZM
corresponds to the transition probability $P_{+}$ in LZT, i.e., $%
D_{n}=P_{+}$. In scheme $\mathbf{B}$, the system starts \
from the center of the avoided crossing, i.e., $\epsilon _{i}=0$,
and evolves to the adiabatic region till far away from the avoided crossing. 
Similarly, there are two regions, an impulse region $[0,\hat{t}]$ and an
adiabatic $(\hat{t},$ $\infty ),$ can be defined. For both schemes, we
directly measure the time-resolved $P_{+}$ of LZT in our experiment and then quantitatively
compare the result with the prediction of KZM.

\noindent\textbf{The time-resolved LZT.} We use superconducting qubits to
investigate the dynamics of the Ising model. Two samples were studied: a
superconducting phase qubit ( denoted as $Q_1$ with $T_1=113$ ns, $T_2^*=93$
ns) \cite{Tan} and a 3D transmon ( denoted as $Q_2$ with $T_1=2.386$ $\mu$s, 
$T_2^*=2.135$ $\mu$s). Here $T_1$ is the energy relaxation time from state $%
|1\rangle$ to state $|0\rangle$, and $T_2^*$ is the
decoherence time including contributions from both relaxation and dephasing. 
Because neither the phase qubit nor the transmon
qubit possess an intrinsic avoided level crossing, we use a 
coherent microwave field to generate an adjustable effective avoided
energy level crossing \cite{Sun}. The position and the
tunneling amplitude ($\Delta )$ of the avoided crossing are
determined by the frequency and amplitude of the microwave field,
respectively. With this flexibility and controllability, instead
of sweeping the flux bias, we chirp the microwave frequency $\omega$ while
keeping the qubit frequency $\omega_{10}$ constant 
to realize LZT (see Methods). The profiles of the control and
measurement pulses are illustrated in Fig. 1b.

In our experiment, scheme $\mathbf{A}$ is a good approximation to LZT from $%
-\infty $ to $+\infty $ because $\epsilon _{i}/2\pi =-200$ MHz and $\Delta
/2\pi =20$ MHz, resulting $|{}\epsilon _{i}|{}/\Delta =10$. The initial
state is prepared in $|1\rangle \approx |{}E_{-}\rangle $ at $\epsilon
_{i}/2\pi =-200$ MHz. 
This is followed immediately by chirping $\epsilon $ to $\epsilon _{f}$ with
a constant speed $v=(\epsilon _{f}-\epsilon _{i})/t_{LZ},$ where $
t_{LZ}$ is the duration of the chirping operation. At the end of $
\epsilon$ chirping, we perform state tomography to determine the density
matrix of the qubit $\rho =\frac{1}{2}(I+\vec{\sigma}\cdot \langle \vec{%
	\sigma}\rangle )$ by measuring the expectation values $\langle \sigma _{x,
	y, z}\rangle $, 
where $\sigma _{x, y, z}$ are the Pauli matrices. 
We then varied $t_{LZ}$ from $1$ ns to $120$ ns, and $\epsilon _{f}/2\pi $
from $-200$ MHz to $+400$ MHz to obtain a complete set of experimental data.
Since the avoided crossing centers at $\epsilon =0$, our experiments
effectively cover the dynamical evolution of the system from $\epsilon \ll
-\Delta $ to $\epsilon \gg \Delta .$

\begin{figure}
	\begin{center}
		\includegraphics[width=.5\textwidth]{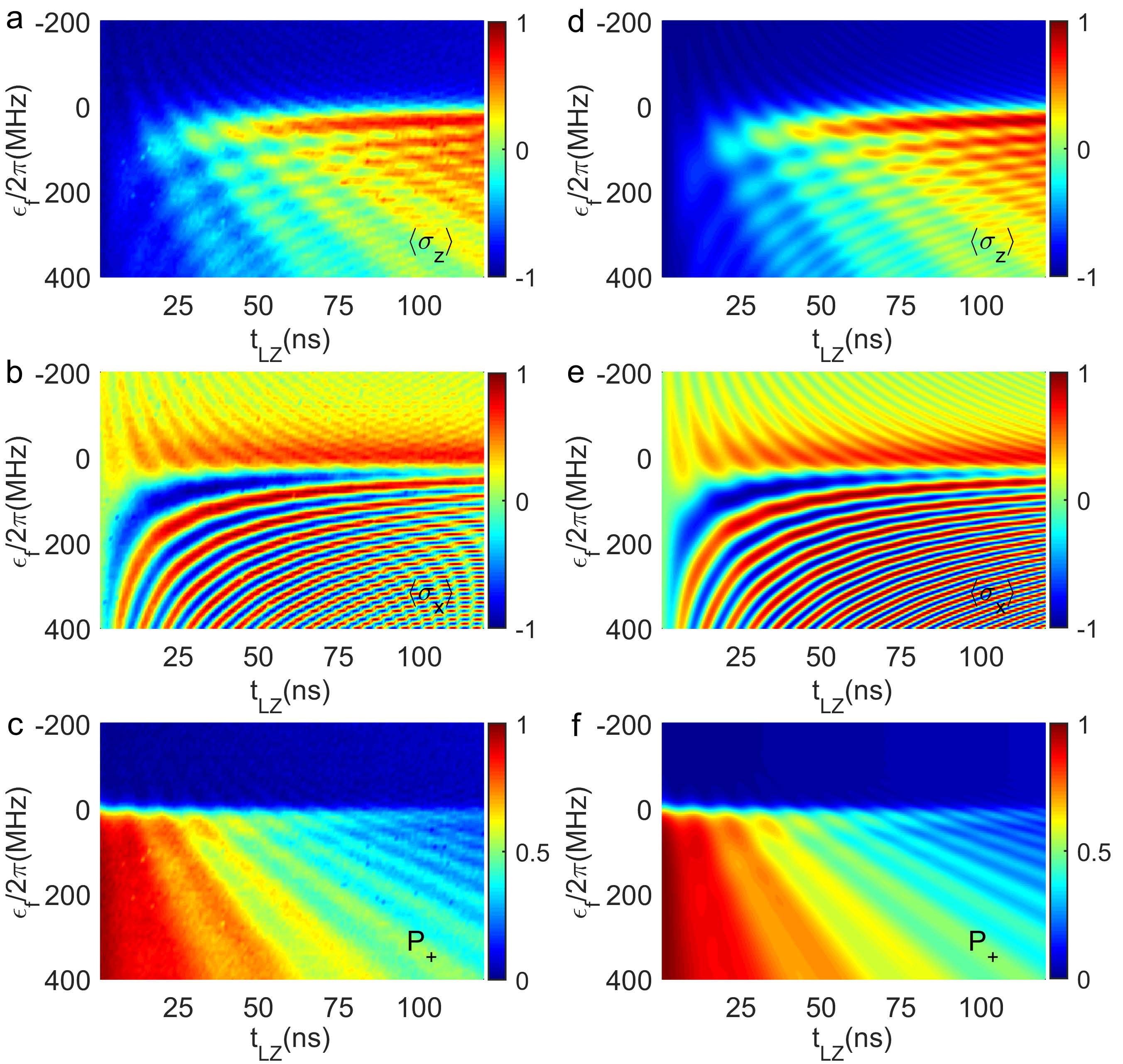}
		\caption{\textbf{The values of $\langle \protect\sigma_{x,z}\rangle$ and $%
				P_{+}$ as a function of $\protect\epsilon _{f}/2\protect\pi$ and $t_{LZ}$.}
			Here $\protect\epsilon _{i}/2\protect\pi=-200$ MHz and $\Delta/2\protect\pi %
			=20$ MHz. The range of $\protect\epsilon _{f}/2\protect\pi$ is from $-200$
			MHz to $400$ MHz. The LZ duration $t_{LZ}$ is from $1$ ns to $120$ ns. 
			\textbf{(a)-(c) ((d)-(f))} are the experimental (numerically simulated)
			results. }
		\label{fig:Graph2}
	\end{center}
\end{figure}

By converting the density matrix to the time-dependent basis $%
|{}E_{\pm}\rangle $, we obtain the 
\begin{equation}
\begin{split}
P_{+}(t) &=\langle E_+(t)|\rho |E_+(t) \rangle \\
&=\frac{1}{2}(1-\langle \sigma _{z}\rangle \cos \theta -\langle \sigma
_{x}\rangle \sin \theta ),  \label{Pplus}
\end{split}%
\end{equation}%
%
which shows that only $\langle \sigma _{z}\rangle $ and $\langle \sigma
_{x}\rangle $ contribute to $P_{+}$. The measured $\langle \sigma
_{z}\rangle $, $\langle \sigma _{x}\rangle $ and $P_{+}$ for qubit $Q_2$ are
plotted in Fig. 2a, b and c, respectively. Shown in Fig. 2d-f are the
results of numerical simulation obtained by solving the master
equations (see Methods), where  important system parameters, such as
the relaxation 
and the dephasing times 
are determined from the pump-decay and the Ramsey fringe measurements. The
good agreement between the experimental and the simulated results indicates
that all essential aspects of our experiment are well controlled and
understood.

\noindent \textbf{The adiabatic and impulse regions.} The time resolved
quantum dynamics of $P_{+}(t)$ described above can be investigated by
measuring $\langle \sigma _{x,z}\rangle $ with nano-second time resolution.
Shown in Fig. 3a are examples of $P_{+}$ for qubit $Q_2$ as a function of
evolution time for various LZT duration time $t_{LZ}$. In order to compare
with AIA, we normalize the evolution time by the freeze-out time $%
\hat{t}=\sqrt{1/\alpha v}.$ The value of $\alpha $ used here is $\pi /2,$
which is the same as that of AIA in this scheme \cite{Damski,damski2006adiabatic}. In
all cases $P_{+}$ changes rapidly near the center of 
the avoided crossing and varies slowly outside the central
region. This is the clear experimental evidence supporting the
physical picture of AIA. Moreover, the boundaries between the
adiabatic and impulse regions are demarcated by $\pm \hat{t}$
with no fitting parameters, confirming the validity of AIA.

In scheme $\mathbf{B}$, we investigate LZT by starting from the
center of the avoided crossing (i.e., $\epsilon _{i}=0).$ The
system is initialized in the lower energy eigenstate 
at $\epsilon _{i}=0$ with a proper resonant $\pi /2$ pulse \cite{NoiseSpectrum, GeoPhase1, GeoPhase2}. 
Then a time sequence similar
to that in scheme $\mathbf{A}$ is applied. Here, $\epsilon
_{f}/2\pi $ ranges from $0$ to $400$ MHz. 
The gap size is fixed at $\Delta /2\pi =20$ MHz, resulting a maximal $%
\epsilon _{f}/\Delta =20$. Shown in Fig. 3b are examples of measured $P_{+}
$ as a function of the evolution time. In this case, $\alpha=\pi/4$
according to Ref. \cite{Damski,damski2006adiabatic}. Similar adiabatic and impulse regions are
observed with the boundary at $t=\hat{t},$ strongly supporting AIA.

\begin{figure}
	\begin{center}
		\includegraphics[width=.5\textwidth]{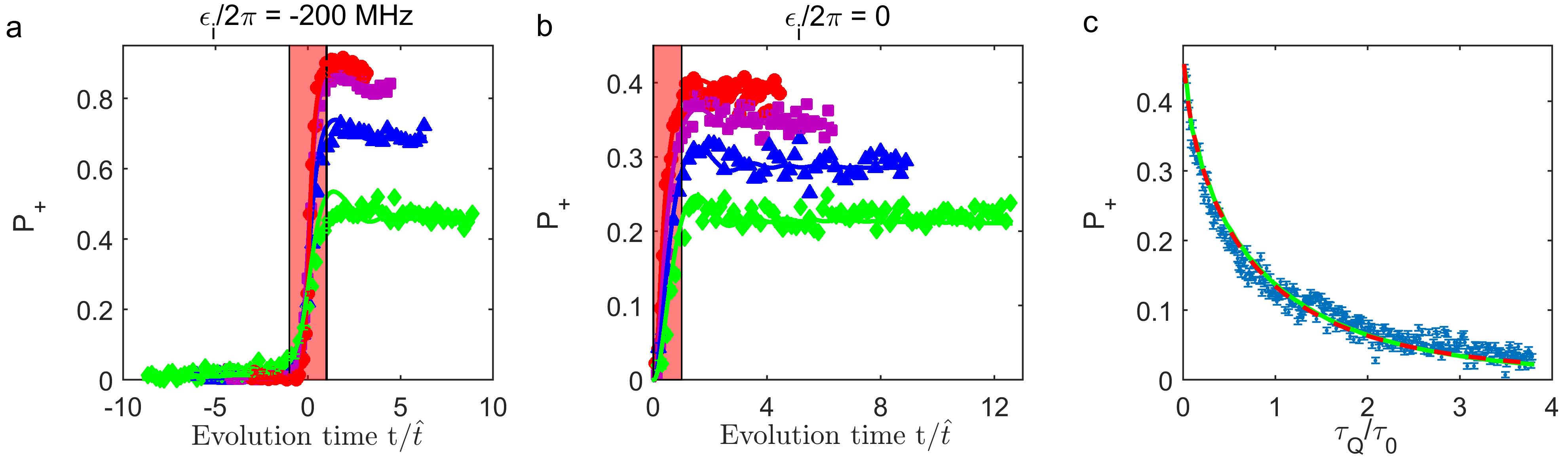}
		\caption{\textbf{Population $P_{+}$ as a function of the normalized time $t/%
				\hat{t}$ and the comparison of $\mathcal{D}_n$ and $P_{+}$. } \textbf{(a)}
			the evolution starting from t = $-\infty$ with $\protect\epsilon _{i}/2\protect%
			\pi =-200$ MHz and $\protect\epsilon _{f}/2 \protect\pi =200$ MHz. 
			\textbf{(b)} the evolution starting from t = $0$ with $\protect\epsilon _{i}/2\protect\pi %
			=0$ and $\protect\epsilon _{f}/2 \protect\pi =400$ MHz. Different
			LZ durations $t_{LZ}=10$ ns (red circle), $20$ ns (magenta square), $40$ ns
			(blue triangle), $80$ ns (green diamond), are used to produce different LZT
			speed. The symbols (solid lines) are experimental (numerical) results. The
			red translucent (clear) regions mark the impulse (adiabatic) regions, while
			the boundary locates on ${\pm \hat{t}}$. The error bars are
			smaller than the sizes of the symbols. \textbf{(c)} The comparison of
			topological defects density $\mathcal{D}_n$ in KZM theory and $P_{+}$ in
			LZT with $\epsilon_i/2\pi=0$. The blue symbols (green solid lines) are the experimental (numerical)
			results. The red dashed line shows the density $\mathcal{D}_{n}$ predicted
			in KZM with $\protect\alpha =0.784$ as the best fit.}
		\label{fig:Graph3}
	\end{center}
\end{figure}

\noindent \textbf{The freeze-out phenomenon.} Another interesting problem is
whether one can observe directly the predicted state freeze-out phenomenon
in the impulse region $[-\hat{t},\hat{t}]$. According to KZM,
although $P_{\pm }$ of the time-dependent basis states $|E_{\pm
}(t)\rangle $ change rapidly in the impulse region, the probability
amplitudes (thus $\langle \overrightarrow{\sigma }\rangle )$ in the
time-independent basis $\{|0\rangle ,|1\rangle \}$ should be frozen 
out. To see whether this is indeed the case, we plot the measured $%
\langle \sigma _{z,x,y}\rangle $ of the  qubit $Q_2$ in Fig. 4a-c (see
Methods). The line represents the freeze-out time $\hat{t}=1/\sqrt{\alpha v}$
is also shown in the plot. Here we use the theoretical KZM value $%
\alpha =\pi /4$ because the total LZT duration is shorter than $40$ ns and
the effect of decoherence may not have significant effects on the
result. It can be seen that $\langle \sigma _{z,x,y}\rangle $ change slowly
in the impulse region, indicating that the state of the qubit is nearly
frozen. In order to compare with the experimental data, we present the
numerically simulated results by solving master equations without adjustable
parameters for the evolution of LZT in Fig. 4d-f. The good agreement between
the simulation and the experimental results supports the observation of the
state freeze-out phenomenon and confirms the validity of AIA.

\begin{figure}
	\begin{center}
		\includegraphics[width=.5\textwidth]{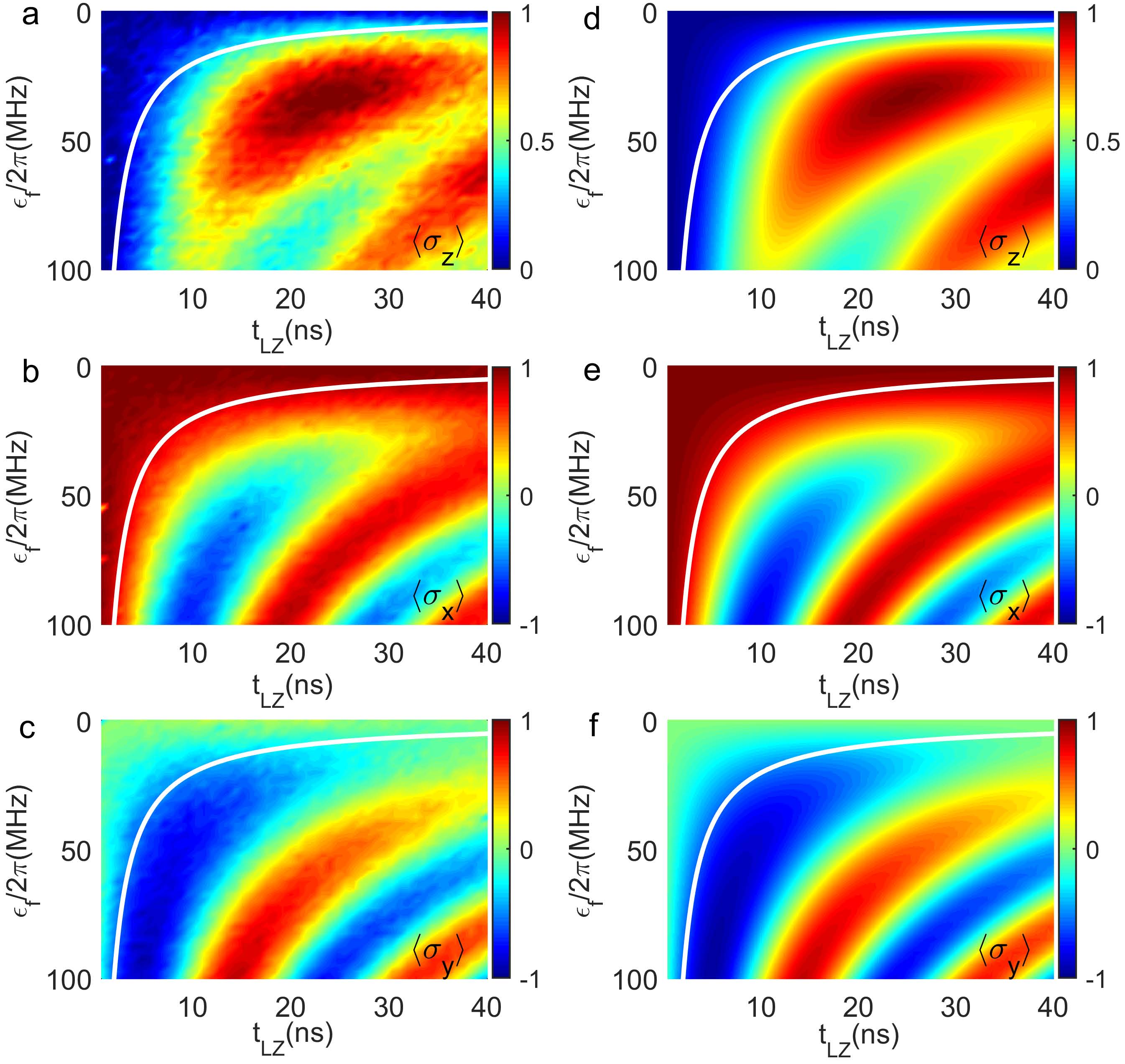}
		\caption{\textbf{State freeze-out phenomena. } \textbf{(a)-(c)} (\textbf{%
				(d)-(f)}) are the experimental observation (numerical simulation) of the
			state freeze-out phenomena of the expectation values $\langle \protect\sigma%
			_{z, x, y} \rangle$ in LZT with $\protect\epsilon _{i}/2\protect\pi =0$. The
			white solid line marks the freeze-out time $\hat{t}$ in KZM with $\protect%
			\alpha =\protect\pi /4$. }
		\label{fig:Graph4}
	\end{center}
\end{figure}

One of the key features of the correspondence between KZM and LZT 
is $\mathcal{D}_{n}\sim P_{+}.$ In Fig. 3c we plot $P_{+}$ as a
function of $t_{LZ}$ for $\epsilon _{f}/2\pi =200$ MHz thus $|\epsilon
_{f}|/\Delta =10$. In order to compare with the theory, $t_{LZ}$ is 
expressed in terms of $\tau _{Q}/\tau _{0}$. It is found that $%
P_{+}$ follows quite well with the behavior of the topological defects
density $\mathcal{D}_{n}$ predicted in Ref. \cite{Damski,damski2006adiabatic}. Here, $\alpha
=0.784$ is obtained from the best fit which is within $0.2\% $ of
KZM predicted value $\pi /4$.  The excellent agreement between the
experimental results of LZT and the theory of KZM  provides strong support
to the conjecture that the dynamics of the Landau-Zener model can be
accurately described in terms of the Kibble-Zurek theory of the topological
defect production in nonequilibrium phase transitions and vice versa.

\noindent \textbf{The scaling law.} We now address the simulation of the
scaling law predicted by KZM for the Ising model. By choosing small
quenching rates $1/\tau _{Q}^{I}$, all the relevant physics described by Eq.
(2) happens in the long wavelength limit $k\ll \pi $. In experiments, we
choose a cutoff $k_{c}/\pi $ and thus $v_{c}/\Delta^2=\chi_{kc}$ to ensure that
LZT probability can be neglected for $|k|>k_{c}$. For each $\tau _{Q}^{I}$,
we choose $N_{k}$ different quasimomentum $k$ equally distributed in $%
[-k_{c},k_{c}]$, and measure the corresponding excitation probability $%
P_{+}(k,\tau _{Q}^{I})$. Then the average density of defects can be
expressed as 
\begin{equation}
\mathcal{N}(\tau _{Q}^{I})=\frac{k_{c}\sum_{k}P_{+}(k,\tau _{Q}^{I})}{\pi
	N_{k}}\sim 1/\sqrt{\tau _{Q}^{I}},  \label{Scaling}
\end{equation}%
where the last equation is given by KZM theory. Stimulated by this
prediction, we plot experimentally measured $\mathcal{N}(\tau _{Q}^{I})$ vs. 
$1/\sqrt{\tau _{Q}^{I}}$  for the qubits $Q_{1}$ (red squares) and $Q_{2}$
(blue squares) in Fig. 5, where $N_{k}=127$ for each $\tau
_{Q}^{I}$, $k_{c}/\pi =0.2$ and $1/\tau _{Q}^{I}<0.01$. A striking feature
is that $\mathcal{N}(\tau _{Q}^{I})$ shows a very good linear relation with $%
1/\sqrt{\tau _{Q}^{I}}.$ By fitting the line to a general linear function $%
\mathcal{N}(\tau _{Q}^{I})=N_{0}+\beta /\sqrt{\tau _{Q}^{I}},$ we obtain the
offset $N_{0}$ and slope $\beta $ as summarized in Table 1. It is
interesting that with the increasing of the decoherence time, the slope
increases while the offset decreases. In order to confirm our observation,
we did numerical simulations using $T_{1}$ and $T_{2}^{\ast }$ of $Q_{1}$, $%
Q_{2},$ and infinite, shown in Fig. 5. Since there is no adjustable
parameter, the agreement between the experimental data and numerical
simulation results are remarkable. When the decoherence time goes to
infinite, the offset tends to zero and the slope  $\beta \simeq 0.106\pm
0.002,$ which is very close to the theory predicted value $1/2\sqrt{2}\pi $ 
\cite{Dziarmaga}. Therefore, it is evident that LZT in qubit exhibits same
scaling behavior of KZM of the Ising model although decoherence will
quantitatively modify the value of some parameters.  

\begin{figure}
	\begin{center}
		\includegraphics[width=.5\textwidth]{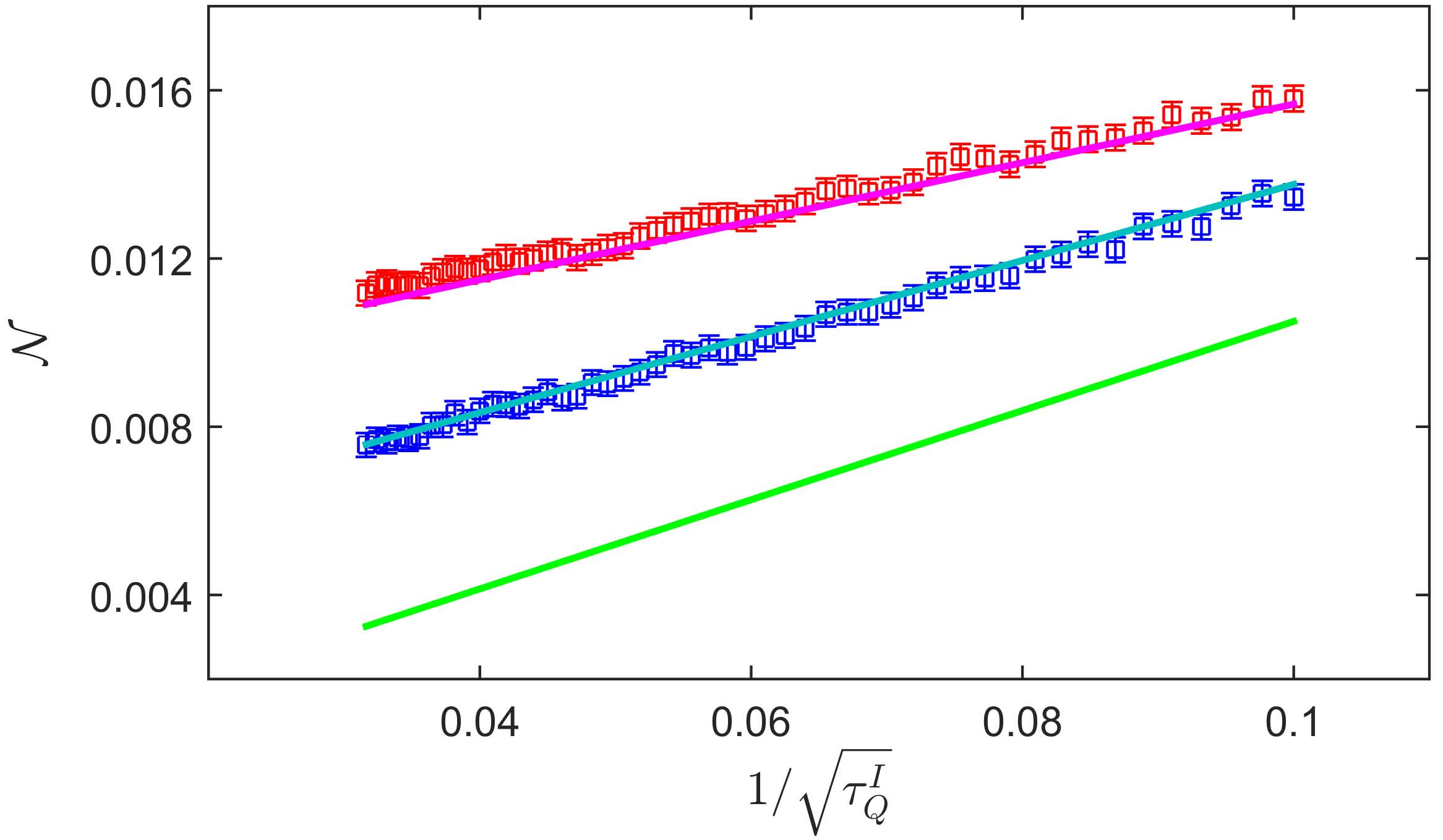}
		\caption{\textbf{The scaling behavior of $\mathcal{N}$ as a function of $1/%
				\protect\sqrt{\protect\tau_Q^I}$.} The red (blue) squares represent the
			experimental data measured in $Q_1$ ($Q_2$). The magenta, blue solid lines
			are the numerical simulation of the master equation with the decoherence of
			the phase qubit and 3D transmon, respectively. The green solid lines are the
			simulated results with infinite $T_1$ and $T_2^*$. }
		\label{fig:Graph5}
	\end{center}
\end{figure}

\noindent\textbf{Discussion}

In our experiment, the hallmark features of the Ising model predicted by 
KZM, such as the existence of the adiabatic and impulse regions, the
freeze-out  phenomenon, and scaling law, were observed. The experimental
observations of the first two are  in good agreement with the theoretical
results because in the  experiment the time of evolution is much less than $%
T_1$ and $T_2^*$ of the qubit  $Q_2$.

We here make several comments on our results of the scaling law. (i) In the
absence of decoherence, our numerical results show that the slope $\beta
\simeq 0.106$, while the theory predicts $\beta \simeq 0.1125$. The
discrepancy between the numerical and the theoretical results is due to the
fact that in the numerical simulation, both $\epsilon _{i}$ and $\epsilon
_{f}$ are finite. the LZ transition happens in a finite range, i.e., $%
\epsilon _{f}/\Delta \simeq 10$. In theory, however, $\beta $ is assuming $%
\epsilon _{f}/\Delta \rightarrow \infty $. To confirm it, we increased $%
\epsilon _{f}/\Delta $ in numerical simulation, and indeed find that $\beta $
approaches $1/2\sqrt{2}\pi $ asymptotically. (ii) The offset $N_{0}$ in the
defect density $\mathcal{N}$ is observed for both qubits in Fig. 5. In addition, the stronger the decoherence, the larger the
offset. It was found in previous studies \cite{Patane,Nalbach} that
decoherence tends to increase the density of defects. As the decoherence
rate increases, more defects are generated in the process of LZ transition.
This is confirmed in our experiments, where it is found that the defect
density $\mathcal{N}$ as well as the offset $N_{0}$ for qubit $Q_{1}$ is
greater than that for qubit $Q_{2}$. If the coherence time increase further,
the offset will gradually approach to zero, as verified by the simulation 3
in Fig. 5. (iii) It is known that the theoretical prediction of Ref. 
\cite{Dziarmaga} is obtained in the continuum limit. So a further question is
whether the finite size effect is important in our experiments. If we
consider a quantum Ising spin chain with a finite length $N$, there will be
energy splitting in the energy spectrum. In this case, if the chirping speed
is too slow compared with the energy splitting caused by the finite size,
then the prediction of Ref. \cite{Dziarmaga} would not be valid. In other
words, the finite size effect sets a lower bound on the chirping speed. To
be more precise, let us consider the energy splitting near zero energy where 
$k=\pm \frac{1}{2}\frac{2\pi }{N}$. Based on Eq. (\ref{LZIsing}), we require 
$\chi_{k=\pm \pi /N}>1$, corresponding to $\sqrt{1/\tau _{Q}^{I}}>\frac{%
	2\pi }{N}.$ In our experiments, we choose $N=N_{k}\pi /k_{c}\simeq 1000$,
which requires $\sqrt{1/\tau _{Q}^{I}}>0.006$, a condition that is very well
satisfied in our experiments. Thus the effect of finite size is negligible.

\begin{table}
	\centering
	\begin{tabular}{lcccc}
		\hline
		Samples & $T_1$ & $T_2^*$ & $\beta$ & $N_0$ \\ \hline
		$Q_1$ & $113$ ns & $93$ ns & $0.068 \pm 0.002$ & $0.0091 \pm 0.0001$ \\ 
		$Q_2$ & $2.386$ $\mu$s & $2.135 $ $\mu$s & $0.088 \pm 0.002$ & $0.0048
		\pm 0.0001$ \\ \hline
		Simu. $1$ & $113$ ns & $93$ ns & $0.070 \pm 0.001$ & $0.0087 \pm 0.0001$ \\ 
		Simu. $2$ & $2.386 $ $\mu$s & $2.135 $ $\mu$s & $0.090 \pm 0.002$ & $0.0048
		\pm 0.0001$ \\ 
		Simu. $3$ & $\infty$ & $\infty$ & $0.106 \pm 0.002$ & $-0.0001 \pm 0.0001$
		\\ \hline
	\end{tabular}
	\caption{ The offset $N_0$ and slope $\protect\beta$ extracted from the
		experimental ($Q_1$, $Q_2$) and simulated results (Simu. 1, Simu. 2, Simu. 3).}
\end{table}

In conclusion, using linear chirps of microwave field and exploring the
correspondence between the KZM of topological defects production and LZT, we
simulated the KZM of the Ising model with a superconducting qubit system.
All important predictions of KZM for the Ising model, such as the existence
of adiabatic and impulse regions, the freeze-out  phenomenon, and especially
the scaling law have been clearly demonstrated. The observed  scaling
behavior in the presence of decoherence sheds new light on the investigation
of the effects of decoherence on KZM of non-equilibrium quantum phase
transitions.

\noindent\textbf{Methods}\newline
\noindent \textbf{Chirp-LZT operation.} In order to perform the LZT, we
chirp the microwave frequency \cite{GeoPhase1, GeoPhase2} instead of
sweeping the flux bias of the qubit. Concretely, with the qubit dc-biased at
a fixed flux $\Phi$, we chirp the microwave frequency from $\omega_i$ to $%
\omega_f$, corresponding to the change of $\epsilon$ from $\epsilon_i$ to $%
\epsilon_f$. Note that in our experiment, the energy difference is $%
\epsilon_{i,f}=\hbar(\omega_{01}-\omega_{i,f})$, with a chirping microwave
frequency $\omega_{i,f}=\omega_{01}+\delta_{\omega_{i,f}}$. Here we assume $%
\hbar=1$. If we set the original microwave frequency as $\omega_{01}$, then
the chirped frequency is $\delta_{\omega_{i,f}}$. Therefore, the
relationship between the chirped microwave frequency $\delta_{\omega_{i,f}}$
and the diabatic energy difference $\epsilon_{i,f}$ is given by $%
\epsilon_{i,f}=-\delta_{\omega_{i,f}}$.

To chirp the microwave frequency, we apply modulation signals from a
Tektronix AWG70002 to the IF (intermediate frequency) ports of a IQ mixer.
Considering the original microwave waveform as $A_r \sin \omega_0 t $,
the modulation signals applied to the I and Q ports of the IQ mixer as $%
\cos \delta_\omega t $ and $\sin \delta_\omega t $, respectively. In this
way, the modulated microwave at the output port of the mixer is $A_r
\sin \omega_0 t   \cos \delta_\omega t +A_r \cos \omega_0 t 
\sin \delta_\omega t =A_r \sin (\omega_0+\delta_\omega) t $, and the
microwave frequency is chirped by tuning $\delta_\omega$. From calibration, 
we find that the power of $\omega_{0}$ tone is at least $50$ dB lower than that of $\omega_0+\delta_\omega$, 
indicating the negligible effect of $\omega_{0}$ on the qubit in the chirp operation.

The Chirp-LZT method provides us several advantages in performing LZT. First
of all, all parameters of the qubit, such as $T_1$, $T_2^*$ and the coupling
strength between the qubit and the external driven filed, are fixed during
the measurements.  Second, the end points of diabatical energy sweep $%
\epsilon_i $ and $\epsilon_f$ would not be  limited by the avoided level
crossings resulting from the coupling between the qubit and microscopic two
level systems (TLSs) usually presented  in superconducting qubits because
the microwave couples much weakly to TLSs than to the qubit. Third, it is
easy to control the chirping velocity and the tunnel splitting $\Delta$ by
controlling the frequency and power of the microwave. In addition, such
method provides a useful tool for systems without natural avoided crossings
in their energy diagram, such as transmons, to perform LZT and other similar
experiments.

\noindent \textbf{Solution of the master equation.} The numerical results
are obtained by solving master equations. The quantum dynamics of the
superconducting qubits is described by the master equations of the time
evolution of the density matrix $\rho$ including the effects of dissipation: 
$\dot {\rho} =\frac{1}{i\hbar}[H \rho]-\Gamma[\rho],$ where $H$ is the
Hamiltonian of the system described by Eq. (3). The second term of this
equation, $\Gamma[\rho]$, describes the effects of decoherence on the
evolution phenomenologically. Setting $\hbar=1$,  the master equation can be
rewritten as

\begin{equation}  \label{MEq}
\left\{ 
\begin{array}{rl}
& \dot{\rho}_{00}=\frac{i}{2}\Delta(\rho_{10}-\rho_{01})+\Gamma_1\rho_{11}
\\ 
& \dot{\rho}_{11}=\frac{i}{2}\Delta(\rho_{01}-\rho_{10})-\Gamma_1\rho_{11}
\\ 
& \dot{\rho}_{01}=\frac{i}{2}[\Delta(\rho_{11}-\rho_{00})+2\epsilon%
\rho_{01}]-\gamma\rho_{01} \\ 
& \dot{\rho}_{10}=\frac{i}{2}[\Delta(\rho_{00}-\rho_{11})-2\epsilon%
\rho_{10}]-\gamma\rho_{10}. 
\end{array}%
\right.
\end{equation}
Here $\Gamma_1\equiv 1/T_1$ is the energy relaxation rate, 
and $\gamma\equiv 1/T_2^*$ is the
decoherence rate.
The relationship between the density matrix and the qubit state expectation
values are given by 
\begin{equation}
\left\{ 
\begin{array}{rl}
& \rho_{00}=(1+\langle \sigma_z \rangle)/2 \\ 
& \rho_{11}=(1-\langle \sigma_z \rangle)/2 \\ 
& Re(\rho_{01})=\langle \sigma_x \rangle/2 \\ 
& Im(\rho_{01})=-\langle \sigma_y \rangle/2. 
\end{array}
\right.
\end{equation}
The numerical simulations in Fig. 2 - 4 are straightforwardly obtained by
solving Eq. (\ref{MEq}). The simulations (solid lines) in Fig. 5 are
obtained by mapping $P_+$ as defined in Eq. (\ref{Pplus}) to the defect
density $\mathcal{N}$. 
By using $v/\Delta^2=\chi_{k}$ for each chirping velocity $v$, we can find the
corresponding momentum $k$ with $\tau_Q^I$ fixed. In this way, we obtain $%
P_+(k,\tau_Q^I)$ for different momentum $k$. Then $\mathcal{N}(\tau_Q^I)$ is
computed from $P_+$ using Eq. (\ref{Scaling}).


\textbf{Acknowledgments}  This work was partly supported by the SKPBR of
China (Grants No. 2011CB922104 and 2011CBA00200), NSFC (Grants No.
91321310, 11274156, 11125417, 11474153, 11474154 and 61521001), NSF of
Jiangsu (Grant No. BK2012013), PAPD, and the PCSIRT (Grant No. IRT1243). S.
Han is supported in part by the U.S. NSF (PHY-1314861).

\textbf{Author contributions} The data were measured by M.G., G.S., Y.Z.,
Y.F., Y.L. and X.T. and analyzed by M.G., X.W., G.S., D.-W.Z., Y.Y., S.-L.Z., S.H. 
and P.W.; H.Y. and D.L. fabricated the sample; The theoretical framework was
developed by X.W. and S.-L.Z.; All authors contributed to the discussion of
the results. Y.Y. and S.H. supervised the experiment and S.-L.Z. supervised
the theory.

\textbf{Additional information}  Correspondence and requests for materials
should be addressed to Y. Y. (email: yuyang@nju.edu.cn) or S.-L.Z. (email:
slzhu@nju.edu.cn) or S.H. (email: han@ku.edu). 

\textbf{Competing financial interests} The authors declare no competing
financial interests.

\end{document}